# Efficient, designable, and broad-bandwidth optical extinction via aspect-ratio-tailored silver nanodisks

E. L. Anquillare,[1,*] O. D. Miller,[2] C. W. Hsu,[1,3] B. G. DeLacy,[4] J. D. Joannopoulos,[1] S. G. Johnson,[2] and M. Soljačić[1]

[1]*Department of Physics, Massachusetts Institute of Technology, Cambridge, MA 02139 USA*
[2]*Department of Mathematics, Massachusetts Institute of Technology, Cambridge, MA 02139 USA*
[3]*Department of Applied Physics, Yale University, New Haven, CT 06520, USA*
[4]*U.S. Army Edgewood Chemical Biological Center, Aberdeen Proving Ground, MD 21010, USA*

*[*eanquill@mit.edu](mailto:eanquill@mit.edu)*

**Abstract:** Subwavelength resonators, ranging from single atoms to metallic nanoparticles, typically exhibit a narrow-bandwidth response to optical excitations. We computationally design and experimentally synthesize tailored distributions of silver nanodisks to extinguish light over broad and varied frequency windows. We show that metallic nanodisks are 2–10x more efficient in absorbing and scattering light than common structures, and can approach fundamental limits to broadband scattering for subwavelength particles. We measure broadband extinction per volume that closely approaches theoretical predictions over three representative visible-range wavelength windows, confirming the high efficiency of nanodisks and demonstrating the collective power of computational design and experimental precision for developing new photonics technologies.

## 1. Introduction

Metallic nanoparticles can scatter and absorb far more light than is directly incident upon them, providing an attractive "plasmonic" [1] platform for enhanced light–matter interactions at optical frequencies. Broad-bandwidth response is a key metric for a wide range of applications including biochemical sensors and immunoassay labels [2], enhanced surface spectroscopy methods [3], colorimetric detection [2], solar-cell textures [2-6] and absorbers [7-9]. However, broad response can be difficult to achieve with nanoparticles, which typically exhibit narrow linewidths [10, 11]. In this paper, we theoretically predict and experimentally demonstrate that disordered collections of metallic nanodisks can provide optimal broad-bandwidth optical extinction (absorption + scattering) for a given material volume, centered around frequency ranges controlled by aspect-ratio-driven wet-synthesis approaches.

While it is well known that the resonant frequencies of nanoparticles can be tuned by altering some combination of the particle sizes [12], shapes [12, 13], and coatings [14], these approaches vary significantly in their efficiencies. Here we show that these nanodisks extinguish 2–10x more visible light (depending on the materials and required bandwidth) than typical structures, for a given material volume. We employ a reproducible wet-synthesis protocol to synthesize such silver nanodisks with aspect-ratio distributions matching computationally optimized designs. Over three representative visible-wavelength windows, $\lambda$ = 400–600nm, 600–800nm, and 400–800nm, the measured per-volume extinction values are greater than 100 $\mu m^{-1}$ over a broad bandwidth—a response that is not theoretically possible via coated-particle [15], dielectric-sphere [16], or multi-channel-resonator [17] geometries. The response is increased at longer wavelengths, reinforcing the important but counterintuitive idea [18, 19] that plasmonic response can increase away from a material's plasma frequency. More generally, our computational and experimental approach is readily adaptable to alternative materials [20] and to applications with an even wider variety of figures of merit.

The optical response of a nanostructure can refer to many linear [21] and nonlinear [22] metrics. Metallic nanoparticles support plasmon resonances [23] that enable such interactions at very subwavelength scales, and recent experimental advances have made it possible to synthesize nanoparticles in a great variety of shapes and sizes [24, 25]. Thus an important design question has emerged: what are the optimal structures and materials for a given application?

Here we consider the problem of broad- and/or tunable-bandwidth absorption and scattering via nanoparticles, and design and synthesize a collection of nanoparticles for optimal broadband extinction. For many such applications, the ideal arrangement is a dilute concentration (which avoids multiple scattering) of particles without coagulation, so that the collective response of the solution is given by the average of the strong individual-particle resonances. (Resonances from multiple particles intentionally placed

at close vicinity [26] or tethered together [27] can be considered single "particles" in this framework.)

The "cross-section" σ of a nanoparticle over a given bandwidth is defined as the power it absorbs and/or scatters divided by the intensity of an incident plane wave. It has units of area and represents a particle's effective optical size, which can be much larger than the particle's physical size due to resonant enhancement [17]. The cross-section itself is unbounded (arbitrarily large particles have arbitrarily large cross-sections), but amount of material is a typical constraint and suggests that the natural figure of merit (FOM) is the cross-section per volume, $\sigma/V$. We optimize the extinction cross-section, $\sigma_{ext}=\sigma_{abs}+\sigma_{scat}$, to encapsulate both absorption and scattering. For a broadband response, one could use a weighted integral over the frequency range of interest, but we find that even for broad bandwidths this choice of FOM leads to single resonances that extinguish light strongly only over a narrow frequency range. To promote enhanced response over many wavelengths, $\lambda$, we instead maximize the worst-case (i.e. least) extinction per volume, as a function of wavelength, over a bandwidth between $\lambda_1$ and $\lambda_2$:

$$\text{FOM} = \min_{\lambda \in [\lambda_1, \lambda_2]} \frac{\sigma_{ext}(\lambda)}{V} \quad (1)$$

We define $\sigma_{ext}(\lambda)$ as the polarization- and angle-averaged extinction to account for the random orientations in a disordered collection of nanoparticles. The FOM has units of inverse-length, representing the inverse thickness of a virtual box that has the volume of the particle and the area of the particle's optical cross-section.

## 2. Theoretical approach and computational design

In theory, it is possible to generate an arbitrarily large $\sigma_{ext}/V$ at a given wavelength with a lossless subwavelength resonator that exhibits a cross-section $\sim\lambda^2$ even as the volume goes to zero. However, the scattering bandwidth must be smaller than the Wheeler–Chu limit [10], which requires $\Delta\lambda \leq 6\pi^2 V_{BS}/\lambda^2$, where $V_{BS}$ is the volume of the object's bounding sphere. For three-dimensional scatterers with fixed geometric aspect ratios, then, the bandwidth must go to zero as the volume does, thereby limiting the integrated response. Lossy nanoparticles exhibit different physics that can lead to greater broadband response. Lossy nanoparticles cannot achieve $\lambda^2$ cross-sections at arbitrarily small sizes; eventually dissipation overtakes radiation and reduces the cross-section [28], imposing strict limits [19] on the peak $\sigma_{ext}/V$. Loss enables the particles to exhibit quasistatic resonances, though, for which the bandwidth is fixed by the material susceptibility $\chi$ [29]. proportional to |Re χ|/Im χ independent of size [30], potentially leading to greater broadband response.

Here we compare four possible approaches to designing particles with large, broadband response. We consider three plasmonic (metallic) approaches: varying particle size (i.e. spheres of different radii), adding coating layers (i.e. metal-on-dielectric spheres), and varying the particle shape (i.e. nanodisks instead of spheres). A fourth approach is the use of wavelength-scale dielectric particles, which avoid the narrow response of metals [16] but also suffer from their lack of subwavelength resonances. We further include a recent analytical limit to the broadband response of any collection of subwavelength particles [31], whose worst-case $\sigma_{ext}/V$ is bounded above by a quantity proportional to the inverse of the average of $[\partial(\varepsilon'\omega)/\partial\omega]/|\chi|^2$ over the desired wavelength range, where $\varepsilon'$ is the real part of the permittivity.

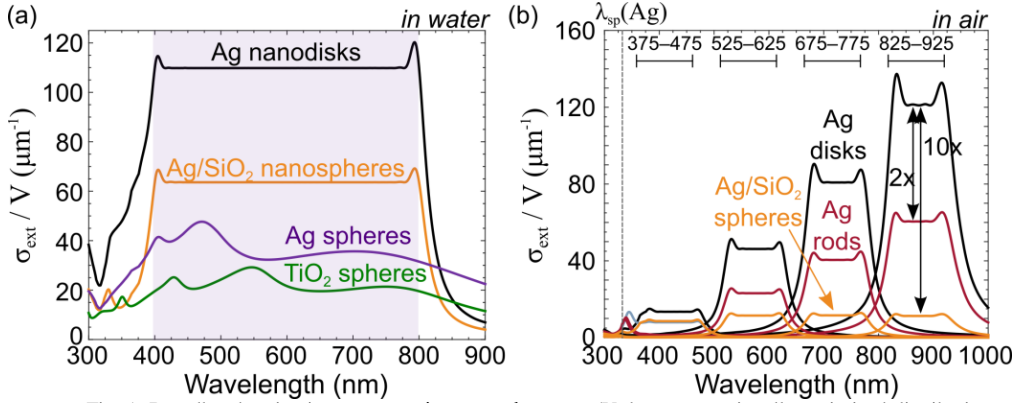

Fig. 1. Broadband extinction cross-section per volume, $\sigma_{ext}/V$, by computationally optimized distributions of nanoparticles (solid lines), alongside theoretical limits to broadband response from subwavelength particles (dashed lines). (a) Over visible wavelengths, tailored distributions of silver nanodisks are better than distributions of more common alternatives, including core-shell ($SiO_2$–Ag) particles and dielectric ($TiO_2$) or metal (Ag) spheres. (b) Across tunable wavelength windows, nanodisks offer significant and increasing enhancements away from silver's bulk plasma wavelength, $\lambda_p(Ag) \approx 324$nm. Detailed theoretical aspect-ratio data is given in Appendix B.

For each particle type described above we consider a large space of ≈1000 structural parameters. The collective extinction per volume is given by the average $\Sigma_i f_i [\sigma_{ext}/V]_i$, where $f_i$ is the volume fraction and $(\sigma_{ext}/V)_i$ the extinction per volume of the $i^{th}$ particle. We use a spherical-wave decomposition [32] to solve for the optical response of the metallic and dielectric spheres over many possible radii. For the nanodisks, single-particle optimizations suggest that quasistatic (size→0) particles are ideal even for broadband response [33], so we use a quasistatic approximation [29] (sometimes referred to as a Gans model [34]) of the particle response. In this case our design space consists of variable aspect ratios (diameter divided by thickness) since they determine the quasistatic response [29] – the overall size does not enter into the computation, since in the quasistatic regime the response is proportional to the volume, which has been normalized out. (One can expect quasistatic response for particles with an effective radius, i.e. equivalent volume, smaller than roughly 30nm [35].) For each type of particle, the degrees of freedom are the volume fractions $\{f_i\}$, which are optimized with an interior-point algorithm [36], a standard linear programming technique [37] that is implemented with the `linprog()` function in MATLAB [38].

Figure 1 compares the response of optimized distributions of silver nanoplates and nanorods, silver-coated silica nanospheres, and dielectric titanium dioxide spheres. Silver was chosen for the metallic material due to its ideal plasmonic properties [18]. Its susceptibility was interpolated from experimental data [39]. (Simulations for higher-quality silver [40], typically describing epitaxial silver, are included Appendix A.) Figure 1(a) compares the optimal collective response for nanoparticles dispersed in water, designed to extinguish light over the 400–800nm range. Although the $TiO_2$ spheres individually exhibit broadband response, their per-volume extinction is relatively low due to the need for the particles to be wavelength-scale to support resonant modes. Conversely, both the core-shell and uncoated (i.e. bare) silver nanodisk structures were optimal at quasistatic sizes, much smaller than the wavelength but still capable of supporting plasmonic resonances. The broad bandwidths of the collective response are achieved through precise engineering of the nanodisk aspect-ratio distribution (ranging from 1.5:1 to 10:1) and of the core-shell filling fraction (ranging from 25% to 95%), respectively. The nanodisks are superior because of their larger polarization currents, which at a single frequency are proportional to $|\chi|^2/\text{Im}\,\chi$, where $\chi$ is the wavelength-dependent susceptibility; in comparison, the peak per-volume extinction in core-shell particles scales [19] only as $|\chi|/\text{Im}\,\chi$, due to the small filling fraction of the metal. The relative bandwidths of such nanoparticles are approximately $|\text{Re } \chi|/\text{Im } \chi$, independent of shape, such that the integrated response of a coated-sphere nanoparticle does not increase with the susceptibility. By contrast, the response of the nanodisks increases as $|\chi|$, which can be on the order of 25 for silver at near-infrared wavelengths. The stark contrast between the two designs is depicted in Fig. 1(b), where we consider four wavelength ranges with 100nm bandwidths. The optimal particles yield nearly identical response around 400nm, where the optimal structure is nearly a bare sphere (since $\text{Re } \chi \approx -3$), but at longer wavelengths the nanodisk response increases significantly, whereas the response of the coated nanospheres remains nearly constant. Included in Fig. 1(b) is the optimal broadband response of nanorods, which have a similar scaling as nanodisks but have roughly half of the per-volume angle-averaged (or orientation-averaged) response, due to their weak coupling to polarizations not parallel to the long axis of the rod [29]. Incident plane waves polarized along the off-axes excite "transverse" modes that have a weak response very close to the flat-surface plasmon frequency—represented by the small bumps near 350nm wavelength in Fig. 1(b). We can also compare to plasmonic particles with multiple aligned resonances ("super-scatterers") [17], not pictured, which for lossy media are designed in [17] to have peak per-volume cross-sections of approximately $16/\lambda$. Such a cross-section could yield at most $\sigma_{ext}/V \approx 40\mu m^{-1}$, for a narrow bandwidth. The nearly wavelength-scale size required to couple to multiple channels diminishes the per-volume response.

### 3. Experimental synthesis and measurement

Having established that tailored-aspect-ratio silver nanoplates are ideal for visible-range extinction, we now discuss their experimental realization. The basis of silver nanoplate formation is the reduction of silver cations to neutral silver atoms that, at a critical concentration, nucleate and grow into nano-sized crystal seeds [41]. The two-dimensional, kinetic growth that follows seed formation has been attributed to several factors [24, 41-46]. Transmission Electron Microscope (TEM) images of our synthesized nanoparticles are shown in Figs. 2(a) and 2(b), respectively, designed for extinction over the 400–600nm and 600–800nm wavelength ranges. A nearly equimolar mixture of the two (54.2% 400-600nm particles and 45.8% 600-800nm particles) exhibited coverage of the entire visible range, $\lambda = 400$–800nm. (Further details on particle synthesis and analysis are given in Appendix C.)

As discussed, the aspect ratio distribution determines the center wavelength and bandwidth of the collective extinction. However, nanoparticle thickness is rarely reported in the literature, let alone aspect-ratio distributions. To evaluate the aspect ratios, we use TEM images of particles flipped on their sides [47] to determine the individual width, thickness, and aspect ratio. By comparing the width distribution of "flipped" particles to the width distribution of all visible particles, we confirm that these particles are representative of the entire population. We find that a slightly modified two-step synthesis based on the method of Li et al. [44] yields particles with moderate aspect ratios ($\approx$2–5), ideal for the 400–600nm range, whereas the single-step reduction synthesis of Metraux et al. [47] more consistently yields the larger aspect ratios ($\approx$5–10) needed for large response in the 600–800nm range.

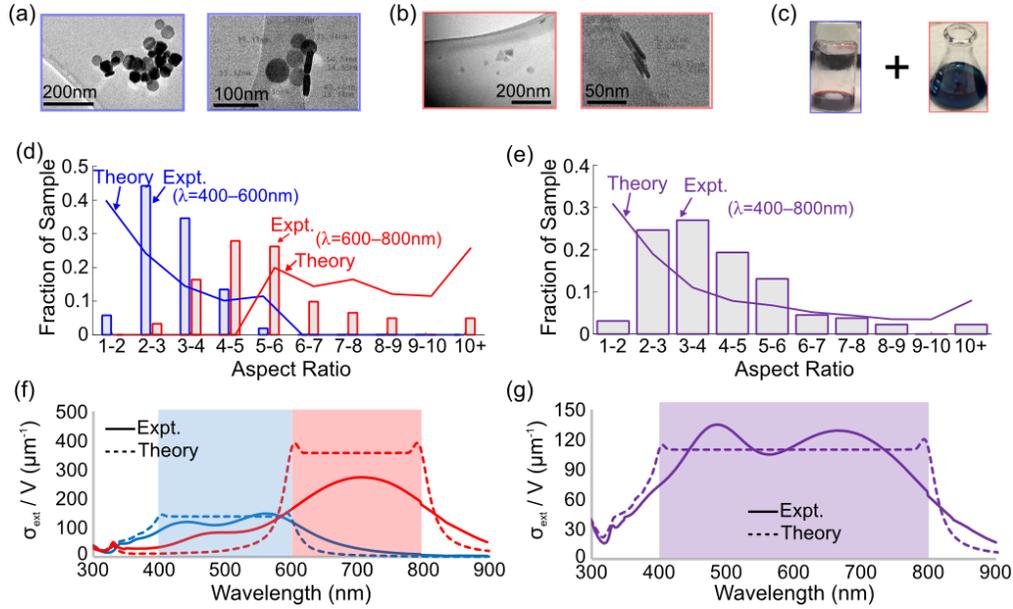

Fig. 2. (a-b) TEM images of nanoparticles synthesized for wavelength ranges (a) λ = 400–600nm (blue) and (b) λ = 600–800nm (red). Particle dimensions range from ≈5–50 nm for their shortest and longest dimensions, respectively. (c) A mixture of the nanoparticles in (a) and (b) enables coverage of λ = 400–800nm (purple). (d-e) Theoretically optimized and experimentally measured aspect ratios, which are on the order of 2-5 for shorter wavelengths (blue) and 3–10 for longer wavelengths (red). (f-g) Experimental measurement (solid) and computational optimization (dashed) of broadband extinction in the three target wavelength ranges. The measured per-volume extinction closely approach the computationally optimized values due to the small particle sizes and nearly matched aspect-ratio distributions.

Figures 2(d) and 2(e) compare the experimentally measured (bars) and computationally optimized (lines) aspect-ratio distributions, for the 400–600nm (blue), 600–800 nm (red), and 400–800nm (purple) ranges. The particles designed to extinguish light at 600–800nm wavelengths had average diameters (or edge lengths, for triangles) of 27nm with average thicknesses of 6nm, while the particles designed for 400–600nm wavelengths had 55nm average diameters and 20nm average thicknesses. Exact particle shape is not crucial; as long as the particles are small enough for their response to be approximately quasistatic, polygonal plates theoretically offer similar response for similar aspect ratios. We find a close correspondence between the theoretically optimal and fabricated aspect ratios, which we attribute to synthesis that promotes lateral growth at different phases in particle formation. In the Metraux et al. method, the silver is reduced in a single step, promoting immediate lateral growth from small seeds and yielding thinner plates with larger aspect ratios. In the Li et al. method, the first step forms larger, spherical particles, to which added silver atoms then attach, creating thicker particles with smaller aspect ratios. Since the particles in the former method produce more consistent thicknesses, they also produce larger variation in aspect ratio distribution.

Coilloidal extinction is then analyzed with a UV-Vis transmission spectrometer. From the molar density of silver reagent and the measured transmission, $\sigma_{ext}/V$ is computed by the Beer-Lambert law, assuming all the silver reagent is converted to nanoplates. The actual $\sigma_{ext}/V$ may be even higher than our estimate because some cationic silver reagent may remain in solution, and solid silver atoms may re-dissolve (e.g. via Ostwald ripening [48]). To ensure there are no effects from multiple scattering or coagulation, we confirmed data collected at different concentrations was consistent.

Figures 2(f) and 2(g) show the measured $\sigma_{ext}/V$ over each target wavelength window: Fig. 2(f) the separate 400-600nm and 600nm-800nm windows, and Fig. 2(g) the entire visible (400-800nm) window. The dashed lines indicate the computationally optimized extinction spectra. One can see that the experimental measurements closely approach the computational optimizations.

The distinguishing feature of increased response as a function of wavelength is demonstrated in Fig. 2(f). A succinct numerical comparison is provided in Table I, with more detailed aspect ratio data in Appendix B. Even the worst-case experimental $\sigma_{ext}/V$, which suffers at the edge of the wavelength windows due to the relatively Gaussian distribution of aspect ratios, reaches 44–64% of the theoretical computations. The average $\sigma_{ext}/V$, more representative of the relative experimental performance, reaches 86%, 67%, and 100% of the optimized response over the 400–600nm, 600–800nm, and 400–800nm windows, respectively. The nanoparticles performing closely to the theoretical designs can be attributed to closely achieving the desired aspect-ratio distributions, especially with the particles designed for the 400–800nm window.

**Table 1.** Comparison of Computationally-Optimized and Experimentally-Measured Nanoplate Properties

| Wavelength range | Comp. optimized | | Expt. measured | | |
|---|---|---|---|---|---|
| | Aspect ratio | $(\sigma_{ext}/V)_{min}$ ($\mu m^{-1}$) | Aspect ratio | $(\sigma_{ext}/V)_{avg}$ ($\mu m^{-1}$) | $(\sigma_{ext}/V)_{min}$ ($\mu m^{-1}$) |
| 400–600nm | 2.7 ± 1.5 | 140 | 3.2 ± 0.8 | 120 | 90 |
| 600–800nm | 8.3 ± 2.1 | 360 | 5.5 ± 1.9 | 240 | 160 |
| 400–800nm | 4.2 ± 3.1 | 110 | 4.3 ± 1.8 | 110 | 60 |

In Appendix D, we use the experimental aspect ratio data to compute the expected extinction spectra for the 400-600nm and 600-800nm wavelength windows and find good agreement with the experimentally measured extinction data, especially considering the relatively small sample sizes. A more robust comparison is given in Appendix E, where we perform a least-squares optimization to compute a theoretical aspect-ratio distribution that would yield the experimental extinction data, and we find that it closely resembles the designed aspect-ratio distribution.

## 4. Discussion and Outlook

Note that the experimental aspect-ratio distribution does not need to perfectly match the computationally optimized distribution; for the 600–800nm window there is a slight mismatch, with smaller aspect ratios than originally designed. This yields the small, undesirable extinction "bump" in the 400–600nm range, but otherwise the collective response is still strongest, and nearly uniform, in the 600–800nm range. We attribute this to the *robustness* of the collective-response approach: because we generate a broad-bandwidth response from a wide array of single-particle features, there is some resilience against random experimental variation. For example, a reduction in one aspect ratio, shifting a resonance to higher or lower frequency, can be compensated by a similar variation in the synthesis of another aspect ratio, with the two variations effectively canceling each other. Such a mechanism is pronounced at the center of a broad bandwidth, where variations in particle resonances on either side of the center frequency can cancel each other, and is less pronounced at the "edge" frequencies, where such compensation is less likely. This feature is seen in all of the experimental extinction data that we present: the "edge" frequencies are harder to control and show greater variation. This variation could be reduced experimentally, with greater control over the exact aspect ratios, but it would also be interesting to explore whether this could be controlled theoretically, e.g. by using robust-optimization techniques to further increase tolerance to synthesis error. We leave this idea for future work.

We expect that the particles synthesized and measured here are dominated by absorption due to their small sizes (there is no scattered power in a quasistatic response). The small sizes were chosen because they appear to be optimal for maximum extinction—bounds on scattered power per volume at a given frequency are four times smaller than equivalent bounds for absorbed power [19]—and thus these particles may be useful for theranostic applications. The computational approach we present here could be extended in a straightforward way to design broadband scattering response, with suppressed absorption, with only a small extension of the particle response function to include radiative losses. Such a particle design could apply to a number of applications; in particular, to scatter light into a nearby active layer and enhance broadband absorption efficiency in a photovoltaic cell [7].

To summarize, we have demonstrated a unified computational and experimental approach for generating efficient and near-uniform broadband response from narrow-band resonators. We have shown that silver nanoplates with moderate aspect ratios are superior to more common structures, with enhancements approaching 10x even at frequencies far from silver's bulk plasma frequency. Our results generalize to other materials and frequencies, and our methodology can be applied to other form factors and figures of merit (for example, [49]). Combining computational design with precise fabrication and synthesis enables the use of typical nanoscale elements to achieve novel emergent behavior.

**A. Comparison of optimal extinction for low- and high-quality silver**

Differences in material quality can lead to significantly different single-frequency response (increased loss reduces resonant enhancement), but they lead to smaller variations in broadband response. In the main text we used data from Palik [39] for our theoretical predictions and optimizations; here we include the (small) increase in extinction that would be possible with the higher-quality silver (typical of epitaxial films) of Johnson and Christy [40].

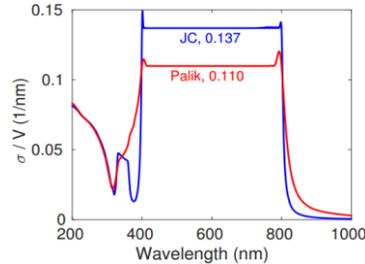

Fig. 3. Optimal extinction per volume (in water), over the 400–800nm wavelength window, for nanoparticles with material susceptibilities described by Palik (red) [39] versus Johnson and Christy (blue, JC) [40].

**B. Optimal aspect-ratio distributions**

Table 2 and Table 3 present more detailed aspect-ratio data to accompany Fig. 1 and Fig. 2 in the main text. Table 2 contains optimal aspect-ratio data for the four extinction windows in Fig. 1(b). Table 3 presents both theoretically optimal and experimentally-measured aspect-ratio data for the optimal distributions in Fig. 2, providing numbers to accompany the bar graphs in Figs. 2(d) and 2(e).

Table 2: Aspect-ratio data for optimal narrow-band response (Fig. 1)

| Narrowband Extinction Window | Aspect Ratio Range (Average) |
|---|---|
| 375-475nm | 2-5 (3) |
| 525-625nm | 7-11 (9) |
| 675-775nm | 14-20 (17) |
| 825-925nm | 23-30 (27) |

Table 3: Percentage Distribution of Aspect Ratios

| Extinction Window | | 1-2 | 2-3 | 3-4 | 4-5 | 5-6 | 6-7 | 7-8 | 8-9 | 9-10 | 10+ |
|---|---|---|---|---|---|---|---|---|---|---|---|
| 400-600nm | Theory | 0.4 | 0.24 | 0.14 | 0.1 | 0.11 | 0 | 0 | 0 | 0 | 0 |
| | Experiment | 0.06 | 0.44 | 0.35 | 0.13 | 0.02 | 0 | 0 | 0 | 0 | 0 |
| 600-800nm | Theory | 0 | 0 | 0 | 0 | 0.2 | 0.14 | 0.16 | 0.12 | 0.11 | .27 |
| | Experiment | 0 | 0.03 | 0.16 | 0.28 | 0.26 | 0.1 | 0.07 | 0.05 | 0 | 0.05 |
| 400-800nm | Theory | 0.31 | 0.19 | 0.11 | 0.08 | 0.07 | 0.05 | 0.04 | 0.04 | 0.03 | 0.08 |
| | Experiment | 0.03 | 0.25 | 0.27 | 0.19 | 0.13 | 0.05 | 0.04 | 0.02 | 0 | 0.02 |

## C. Nanoparticle synthesis and analysis

All chemicals were purchased from Sigma Aldrich or Alfa Aesar and not further purified. All water was deionized via a Millipore Milli-DI system.

Red-wavelength-extinguishing (600-800nm) nanoparticles were synthesized based on the method first presented by Metraux and Mirkin. [47] 25ml of 0.1mM silver nitrate solution was rapidly stirred at room temperature under ambient conditions with 1.5ml 30mM trisodium citrate and 1.5ml 0.7mM Polyvinylpyrrilidone (M.W. 29k). 60µl of 29-32% by weight hydrogen peroxide solution was added to the vial, after which it was coated in tinfoil to ensure a dark reaction environment. Finally, 250µl of 100mM aqueous sodium borohydride solution was added drop-wise via syringe while stirring. After one to two hours, the solution changed color from clear to dark peacock blue, indicating nanoparticle formation. [47]

Blue-wavelength-extinguishing (400-600nm) particles were synthesized by slightly modifying a method first presented by Li et al. [44] 80µl of 1.875M hydrogen peroxide solution and 0.58ml of 0.01M silver nitrate solution were added while stirring to 23.54ml of $2.5 \times 10^{-4}$M trisodium citrate solution. While still stirring in ambient conditions, 600µl of 0.1M aqueous sodium borohydride solution was added drop-wise via syringe. At this point, the solution rapidly turned yellow, indicating the formation of small nanoparticle seeds. 400µl of this solution was then re-dispersed in 1.6ml water and subject to continued stirring. Trisodium citrate (100µl, 0.075M) and Ascorbic Acid (200µl, 0.1M) solutions were then added. Finally, 100µl of 0.1M silver nitrate solution was slowly and steadily added drop-wise over the course of ten minutes. Silver addition immediately changed the color of the solution from yellow to opaque, reddish purple, indicating further nanoparticle growth. [44]

It is important to note that in the latter seed-mediated method, the addition rate of the second dose of silver can have a critical effect on the particle shape, and ergo, the optical response. Adding the silver reagent all at once yielded rounded, irregular platelets with a strong peak at 468nm and shoulder at 387nm. Drop-wise addition at a rate of one drop per minute over the course of six to ten minutes yielded more angular, delineated, geometric nanoparticles with a broader absorption and two smaller peaks. The first peak consistently appeared around 425nm but the location of the second peak varied from 500nm to 600nm, likely due to imprecise quantification of the volume of the individually-added drops. When ultimately seeking to extinguish over certain wavelength windows, rapid addition was better for the 400-600nm particles later mixed to form the 400-800nm window while slow addition was better for the 400-600nm window alone. The particles synthesized with different addition rates had only mildly differing aspect ratios, and their aspect ratios were still both larger when compared to the 600-800nm extinguishing nanoplates. The mixture of nanoparticles used to extinguish over the 400-800nm visible region contained (by mol) 54.2% of the 400-600nm nanoplates and 45.8% of the 600-800nm nanoplates.

It is also important to note that the per-volume extinction coefficients of the particles were evaluated at varying colloid concentrations and yielded results within instrumental error of the original values with no relationship emerging between concentration and extinction value. This affirms that the optical responses achieved indeed represent dispersed systems not affected by interparticle interactions.

Nanoparticles were sized by deposition on a copper grid with a holey carbon formvar film and analyzed using a FEI Tecnai G2 Spirit TWIN transmission electron microscope. Nanoparticle solutions were analyzed using a Cary 500i UV-Vis transmission spectrometer.

### D. Extinction reconstruction

Here we compare the measured extinction to the extinction theoretically predicted from the experimentally measured aspect ratios. The experimental aspect-ratio data is input into scuff-em, an open-source implementation [50] of the boundary element method [51]. We can thus construct the expected extinction and compare to the actual measured extinction values. Figure 2 compares the two curves. The shape of the two curves do not exactly match, but the small sample size for the aspect-ratio data makes the theoretical reconstruction highly sensitive to each data point. More generally, the tunable-window effect is clearly visible in both curves.

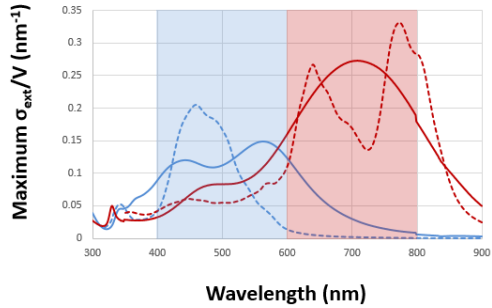

Fig. 4 Comparison of experimentally measured $\sigma_{ext}/V$ spectra (bold) to a theoretical reconstruction (dashed) of the expected spectra given the experimentally measured aspect ratios, for two wavelength windows (red and blue).

### E. Aspect-ratio reconstruction

In this section we perform a comparison between experiment and theory that is complementary to that of the previous section. Here we take the measured extinction data as given and perform a least-squares optimization to find the aspect-ratio distribution that would produce such a spectrum (in a quasistatic approximation). Figure 5(a) shows the theoretically reconstructed spectrum, while Fig. 5(b) shows the aspect-ratio distribution that would produce the theoretical curve (blue) in Fig. 5(a). One can see that the optimal aspect ratios in Fig. 5(b) are concentrated in the range from one to six, just as is seen in the optimal design and experimental measurement in Fig. 2(d).

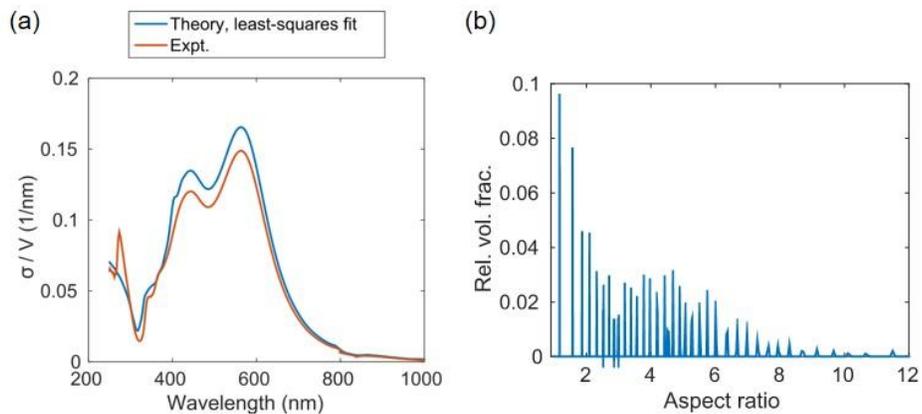

Fig. 5. (a) Reconstruction (blue) of experimental extinction (red) curve via least-squares optimization of the aspect-ratio distribution. (b) Aspect-ratio distribution that produces the theoretical curve in (a) for small, quasistatic silver nanoparticles.

**Acknowledgments**

We thank Yong Zhang & Tim McClure for instrumentation assistance. This work was supported by the U. S. Army Research Laboratory and the U. S. Army Research Office through the Institute for Soldier Nanotechnologies under contract number W911NF-13-D-0001). It was also supported in part by MRSEC Program of the National Science Foundation under contract DMR-1419807 and the MIT Deshpande Center for Technological Innovation.